\documentclass[conference,a4paper]{APSIPA2019}
\usepackage{multirow}
\usepackage[dvips]{graphicx}
\usepackage{amsmath}
\usepackage[psamsfonts]{amssymb}
\usepackage{amsxtra}
\usepackage{threeparttable}

\begin{document}

\title{AP19-OLR Challenge: Three Tasks and Their Baselines}

\author{%
\authorblockN{%
Zhiyuan Tang\authorrefmark{2},
Dong Wang\authorrefmark{2}\authorrefmark{3}$^*$ and
Liming Song\authorrefmark{4}
}
\authorblockA{%
\authorrefmark{2}
Center for Speech and Language Technologies, Tsinghua University\\
}
\authorblockA{%
\authorrefmark{3}
Beijing National Research Center for Information Science and Technology\\
\authorrefmark{4}
SpeechOcean\\
Corresponding email: wangdong99@mails.tsinghua.edu.cn
}
}

\maketitle
\thispagestyle{empty}

\begin{abstract}
  This paper introduces the fourth oriental language recognition (OLR) challenge AP19-OLR,
  including the data profile, the tasks and the evaluation principles.
  The OLR challenge has been held successfully for three consecutive years,
  along with APSIPA Annual Summit and Conference (APSIPA ASC).
  The challenge this year still focuses on practical and challenging tasks, precisely
  (1) short-utterance LID,
  (2) cross-channel LID
  and (3) zero-resource LID.
  
  The event this year includes more languages and more real-life data provided 
  by SpeechOcean and the NSFC M2ASR project.
  All the data is free for participants.
  Recipes for x-vector system and back-end evaluation are also conducted 
  as baselines for the three tasks.
  The participants can refer to these online-published recipes to deploy LID systems for convenience.
  We report the baseline results on the three tasks and
  demonstrate that the three tasks are worth some efforts to achieve better performance.
  
  \end{abstract}
  
  \section{Introduction}

  There are thousands of languages around the world grouping many language families,
  such as the oriental language families which often include 
  Austroasiatic languages (e.g.,Vietnamese, Cambodia)~\cite{sidwell201114},
  Tai-Kadai languages (e.g., Thai, Lao), Hmong-Mien languages (e.g., some dialects in south China), Sino-Tibetan languages (e.g., Chinese Mandarin), Altaic languages (e.g., Korea, Japanese) and Indo-European languages (e.g., Russian)~\cite{ramsey1987languages,shibatani1990languages,comrie1996russian}.
  With so many languages, the development of communication technology and
  movement of worldwide population make multilingual phenomena more and more common,
  and in turn, more advanced speech technologies have been developed to
  further boost the communication in multilingual environment,
  e.g., instant and simultaneous interpretation with machine.

  The language identification (LID) technology plays a great role in the development of
  multilingual interaction between human and machine,
  and it is often located at the front end of other speech processing systems, mostly speech recognition (ASR).
  To better meet the needs of multilingual ASR,
  the building of LID system may encounter many difficult issues, such as high real-time requirement, 
  cross-channel speech signals and very noisy background.


  Considering the languages for which we build the LID system,
  there may also exist a huge difference in linguistic resources between two different languages,
  such as expert knowledge about the language and amounts of digital resources for machine learning.
  Some languages spoken by decreasing number of population may even face the risk of extinction.
  That requires better language technologies to process these low-resource or even zero-resource languages,
  including spoken language identification technology.
  Different languages also interact and influence each other,
  leading to complicated linguistic evolution and lots of research~\cite{huang2013cross,fer2015multilingual,wang2015transfer}.

  The oriental language recognition (OLR) challenge is organized annually,
  aiming at improving the research on multilingual phenomena and
  advancing the development of language recognition technologies.
  The challenge has been conducted three times since 2016,
  namely AP16-OLR~\cite{wang2016ap16}, AP17-OLR~\cite{tang2017ap17} and AP18-OLR~\cite{tang2018ap18},
  each attracting dozens of teams around the world.

  AP18-OLR involved $10$ languages and focused on three challenging tasks: 
  (1) short-utterance ($1$ second) LID, which was inherited from AP17-OLR;
  (2) LID for confusing language pairs; (3) open-set LID where the test data
  involved unknown interference languages.
  In the first task, the system submitted by the XMUspeech team achieved the best performance
  ($C_{avg}$=$0.0462$, EER\%=$4.59$).
  In the second and third tasks, the systems submitted by the NetEase AI-Speech team achieved the best performance
  with $C_{avg}$=$0.0032$, EER\%=$0.33$ and $C_{avg}$=$0.0119$, EER\%=$3.16$ respectively.
  From these results, one can see that for the short-utterance condition, the task remains
  challenging.  More details about the past three challenges can be found
  on the challenge website.\footnote{http://olr.cslt.org}

  Based on the experience of the last three challenges and the calling from industrial application, 
  we propose the fourth OLR challenge.
  This new challenge, denoted by AP19-OLR, will be hosted by APSIPA ASC 2019.
  It involves more languages and focuses on more practical and challenging tasks:
  (1) short-utterance ($1$ second) LID, as in the past two challenges,
  (2) cross-channel LID, which reveals the real-life demand of speech technology such as machine interpretation,
  and (3) zero-resource LID, where no resources are provided for training before inference,
  but only several utterances of each language are provided for language reference.

  In the rest of the paper, we will present the data profile and the evaluation plan of the AP19-OLR challenge. To
  assist participants to build their own submissions, baseline recipes are constructed based on the x-vector system.
  The Kaldi recipes of these baselines can be downloaded from the challenge website.
  
  \begin{table*}[htb]
  \begin{center}
  \caption{AP16-OL7 and AP17-OL3 Data Profile}
  \label{tab:ol10}
  \begin{tabular}{|l|l|c|c|c|c|c|c|c|}
   \hline
  \multicolumn{3}{|c|}{\textbf{AP16-OL7}} & \multicolumn{3}{c|}{AP16-OL7-train/dev}  & \multicolumn{3}{c|}{AP16-OL7-test}\\
  \hline
  Code & Description & Channel & No. of Speakers & Utt./Spk. & Total Utt. & No. of Speakers & Utt./Spk. & Total Utt. \\
  \hline
  ct-cn & Cantonese in China Mainland and Hongkong & Mobile & 24 & 320 & 7559 & 6 & 300 & 1800 \\
  \hline
  zh-cn & Mandarin in China & Mobile & 24 & 300 & 7198        & 6 & 300 & 1800 \\
  \hline
  id-id & Indonesian in Indonesia &  Mobile & 24 & 320 & 7671 & 6 & 300 & 1800 \\
  \hline
  ja-jp & Japanese in Japan & Mobile & 24 & 320 & 7662        & 6 & 300 & 1800 \\
  \hline
  ru-ru & Russian in Russia & Mobile & 24 & 300 & 7190        & 6 & 300 & 1800 \\
  \hline
  ko-kr & Korean in Korea & Mobile & 24 & 300 & 7196          & 6 & 300 & 1800 \\
  \hline
  vi-vn & Vietnamese in Vietnam & Mobile & 24 & 300 & 7200    & 6 & 300 & 1800 \\
   \hline
  \hline
   \multicolumn{3}{|c|}{\textbf{AP17-OL3}} & \multicolumn{3}{c|}{AP17-OL3-train/dev}  & \multicolumn{3}{c|}{AP17-OL3-test}\\
  \hline
  Code & Description & Channel & No. of Speakers & Utt./Spk. & Total Utt. & No. of Speakers & Utt./Spk. & Total Utt. \\
  \hline
  ka-cn & Kazakh in China & Mobile & 86 & 50  & 4200 &      86 &  20  & 1800 \\
  \hline
  ti-cn & Tibetan in China & Mobile & 34 & 330   & 11100 &    34 & 50  & 1800 \\
  \hline
  uy-id & Uyghur in China &  Mobile & 353 & 20   & 5800 &    353 & 5  & 1800 \\
   \hline
  \end{tabular}
  \begin{tablenotes}
  \item[a] Male and Female speakers are balanced.
  \item[b] The number of total utterances might be slightly smaller than expected, due to the quality check.
  \end{tablenotes}
  \end{center}
  \end{table*}

  \section{Database profile}

  Participants of AP19-OLR can request the following datasets for system construction.
  All these data can be used to train their submission systems.
  
  \begin{itemize}
  \item AP16-OL7: The standard database for AP16-OLR, including AP16-OL7-train, AP16-OL7-dev and AP16-OL7-test.
  \item AP17-OL3: A dataset provided by the M2ASR project, involving three new languages. It contains AP17-OL3-train and AP17-OL3-dev.
  \item AP17-OLR-test: The standard test set for AP17-OLR. It contains AP17-OL7-test and AP17-OL3-test.
  \item AP18-OLR-test: The standard test set for AP18-OLR. It contains AP18-OL7-test and AP18-OL3-test.
  \item THCHS30:  The THCHS30 database (plus the accompanied resources) published by CSLT, Tsinghua University~\cite{wang2015thchs}.
  \end{itemize}
  
  
  Besides the speech signals, the AP16-OL7 and AP17-OL3 databases also provide lexicons of all the 10 languages, as well
  as the transcriptions of all the training utterances. These resources allow training acoustic-based or phonetic-based
  language recognition systems. Training phone-based speech recognition systems is also possible, though
  large vocabulary recognition systems are not well supported, due to the lack of large-scale language models.

  A test dataset AP19-OLR-test will be provided at the date of result submission,
  which includes three parts corresponding to the three LID tasks.


  \subsection{AP16-OL7}

  The AP16-OL7 database was originally created by Speechocean, targeting for various speech processing tasks.
  It was provided as the standard training and test data in AP16-OLR.
  The entire database involves 7 datasets, each in a particular language. The seven languages are:
  Mandarin, Cantonese, Indonesian, Japanese, Russian, Korean and Vietnamese.
  The data volume for each language is about $10$ hours of speech signals recorded in
  reading style. The signals were
  recorded by mobile phones, with a sampling rate of $16$ kHz  and a sample size of $16$ bits.
  
  For Mandarin, Cantonese, Vietnamese and Indonesia, the recording was conducted in a quiet environment.
  As for Russian, Korean and Japanese, there are $2$ recording sessions for each speaker: the first session
  was recorded in a quiet environment and the second was recorded in a noisy environment.
  The basic information of the AP16-OL7 database is presented in Table~\ref{tab:ol10},
  and the details of the database can be found in the challenge website or the
  description paper~\cite{wang2016ap16}.

  \subsection{AP17-OL7-test}
  
  The AP17-OL7 database is a dataset provided by SpeechOcean. This dataset contains 7 languages as in AP16-OL7,
  each containing $1800$ utterances. The recording conditions are the same as AP16-OL7. This database is used as
  part of the test set for the AP17-OLR challenge.
  
  \subsection{AP17-OL3}
  
  The AP17-OL3 database contains 3 languages: Kazakh, Tibetan and Uyghur, all are minority languages in China.
  This database is part of the Multilingual Minorlingual Automatic Speech Recognition (M2ASR) project, which is
  supported by the National Natural Science Foundation of China (NSFC). The project is a three-party collaboration, including Tsinghua University,
  the Northwest National University, and Xinjiang University~\cite{wangm2asr}. The aim of this project is to construct speech recognition systems for five minor languages in China (Kazakh, Kirgiz, Mongolia, Tibetan and Uyghur). However, our ambition is beyond that scope: we hope
  to construct a full set of linguistic and speech resources and tools for the five languages, and make them open and free for
  research purposes. We call this the M2ASR Free Data Program. All the data resources, including the tools published in this paper, are released on the web site of the project.\footnote{http://m2asr.cslt.org}
  
  The sentences of each language in AP17-OL3 are randomly selected from the original M2ASR corpus.
  The data volume for each language in AP17-OL3 is about $10$ hours of speech signals
  recorded in reading style.
  The signals were recorded by mobile phones,
  with a sampling rate of $16$ kHz and a sample size of $16$ bits.
  We selected $1800$ utterances for each language as the development set (AP17-OL3-dev), and the rest is used as the
  training set (AP17-OL3-train). The test set of each language involves $1800$ utterances, and is provided separately
  and denoted by AP17-OL3-test.
  Compared to AP16-OL7, AP17-OL3 contains much more variations in terms of recording conditions and
  the number of speakers, which may inevitably  increase the difficulty of the challenge task.
  The information of the AP17-OL3 database is summarized in Table~\ref{tab:ol10}.

  \subsection{AP18-OLR-test}
  The AP18-OLR-test database is the standard test set for AP18-OLR,
  which contains AP18-OL7-test and AP18-OL3-test.
  Like the AP17-OL7-test database,
  AP18-OL7-test contains the same target $7$ languages, each containing $1800$ utterances,
  while AP18-OL7-test also contains utterances
  from several interference languages.
  The recording conditions are the same as AP17-OL7-test.
  Like the AP17-OL3-test database,
  AP18-OL3-test contains the same $3$ languages, each containing $1800$ utterances.
  The recording conditions are also the same as AP17-OL7-test.


  \subsection{AP19-OLR-test}
  The AP19-OLR-test database is the standard test set for AP19-OLR,
  which includes 3 parts responding to the 3 LID tasks respectively, precisely
  AP19-OLR-short, AP19-OLR-channel and AP19-OLR-zero.

  \begin{itemize}
    \item AP19-OLR-short: This subset is designed for the short-utterance LID task,
    which contains the ten target languages as in AP18-OLR-test and each language has $1800$ utterances.
    \item AP19-OLR-channel: This subset is designed for the cross-channel LID task,
    which contains six of the ten target languages as in AP18-OLR-test, but was recorded in wild environment.
    The six languages are Tibetan, Uyghur, Japanese, Russian, Vietnamese and Mandarin.
    Each language has $1800$ utterances.
    \item AP19-OLR-zero: This subset is designed for the zero-resource LID task.
    The three languages are not in the ten traditional languages,
    but other resource-limited languages, namely Catalan, Greek and Telugu.
    Each language has $10$ utterances for reference and $1800$ for identification test.
  \end{itemize}

  To help the participants develop systems against the three tasks, 
  development set AP19-OLR-dev is also provided.
  Specifically, for task 1, the short-utterance test set from AP18-OLR-test
  can be reused.
  For task 2 and 3, a new smaller development set is provided respectively,
  while the three target languages in the third development set are different 
  from those in the final test set.

  \section{AP19-OLR challenge}
  
  The evaluation plan of AP19-OLR keeps mostly the same
  as in AP18-OLR, except some modification for the new
  challenge tasks.
  
  Following the definition of NIST LRE15~\cite{lre15}, the task of the LID challenge is defined
  as follows: Given  a  segment  of  speech  and  a  language  hypothesis (i.e.,  a  target
  language  of  interest  to  be  detected),  the  task  is  to decide  whether  that
  target  language  was  in  fact  spoken  in  the given segment (yes or no), based on an
  automated analysis of the data contained in the segment.
  The evaluation plan mostly follows the principles of NIST LRE15.
  
  
  
  The AP19-OLR challenge includes three tasks as follows:
  
  \begin{itemize}
  \item Task 1: Short-utterance LID is a close-set identification task,
     which means the language of each utterance is among the known traditional $10$ target languages.
     The utterances are as short as $1$ second.
  \item Task 2: Cross-channel LID, where test data in different channels for the known $10$ target languages
      will be provided.
  \item Task 3: Zero-resource LID, where no resources are provided for training before inference,
      but several reference utterances are provided for each language.
  \end{itemize}

  \subsection{System input/output}
  
  The input to the LID system is a set of speech segments in unknown languages.
  For task 1 and task 2, those speech segments are within
  the $10$ known target languages.
  For task 3, the target languages of the speech segments are the same as the reference utterances.
  The task of the LID system is to determine
  the confidence that a language is contained in a speech segment. More specifically,
  for each speech segment, the LID system outputs a score vector $<\ell_1, \ell_2, ..., \ell_{10}>$,
  where $\ell_i$ represents the confidence that language $i$ is spoken in the speech segment.
  The scores should be comparable across languages and segments.
  This is consistent with
  the principles of LRE15, but differs from that of LRE09~\cite{lre09} where an explicit decision
  is required for each trial.
  
  In summary, the output of an OLR submission will be a text file, where each line contains
  a speech segment plus a score vector for this segment, e.g.,
  
  \vspace{0.5cm}
  \begin{tabular}{ccccccccc}
          & lang$_1$   & lang$_2$   & ... & lang$_9$  & lang$_{10}$\\
  seg$_1$ & 0.5  & -0.2 &  ...& -0.3 & 0.1    \\
  seg$_2$ & -0.1 & -0.3 &  ...& 0.5 & 0.3    \\
  ...   &      &     &  ... &      &
  \end{tabular}

  \subsection{Test condition}
  
  
  \begin{itemize}
  \item No additional training materials. The only resources that are allowed to use are:
        AP16-OL7, AP17-OL3, AP17-OLR-test, AP18-OLR-test, and THCHS30.
  \item All the trials should be processed. Scores of lost trials will be interpreted as -$\inf$.
  \item The speech segments in each task
        should be processed independently, and each test segment in a group should be processed
        independently too. Knowledge from other test segments is not allowed to use (e.g.,
        score distribution of all the test segments).
  \item Information of speakers is not allowed to use.
  \item Listening to any speech segments is not allowed.
  \end{itemize}

  \subsection{Evaluation metrics}
  
  As in LRE15, the AP19-OLR challenge chooses $C_{avg}$ as the principle evaluation metric.
  First define the pair-wise loss that composes the missing and
  false alarm probabilities for a particular target/non-target language pair:
  
  \[
  C(L_t, L_n)=P_{Target} P_{Miss}(L_t) + (1-P_{Target}) P_{FA}(L_t, L_n)
  \]
  
  \noindent where $L_t$ and $L_n$ are the target and non-target languages, respectively; $P_{Miss}$ and
  $P_{FA}$ are the missing and false alarm probabilities, respectively. $P_{target}$ is the prior
  probability for the target language, which is set to $0.5$ in the evaluation. Then the principle metric
  $C_{avg}$ is defined as the average of the above pair-wise performance:
  
  
  \[
   C_{avg} = \frac{1}{N} \sum_{L_t} \left\{
  \begin{aligned}
    & \ P_{Target} \cdot P_{Miss}(L_t) \\
    &  + \sum_{L_n}\ P_{Non-Target} \cdot P_{FA}(L_t, L_n)\
  \end{aligned}
  \right\}
  \]

  \noindent where $N$ is the number of languages, and $P_{Non-Target}$ = $(1-P_{Target}) / (N -1 )$.
  We have provided the evaluation script for system development.

  \section{Baseline systems}

  We construct the baseline systems for the three tasks respectively.
  All the experiments are conducted with Kaldi~\cite{povey2011kaldi}.
  The purpose of these experiments is to present a reference for the participants, rather than a competitive submission.
  The recipes can be downloaded from the website of the challenge.

  \subsection{X-vector system}

  We use the x-vector system as described in \cite{snyder2018x,snyder2018spoken}.
  The raw feature of the system is $40$-dimensional filterbanks.
  The energy VAD is used to filter out nonspeech frames.
  The network configuration is outlined in Table~\ref{tab:xvect} as shown in \cite{snyder2018x}.
  The DNN is trained to classify the $N$ languages in the training data. 
  After training, embeddings called `x-vectors' are extracted from the affine component of
  layer $segment6$. Excluding the $softmax$ output layer and $segment7$
  there is a total of $4.2$ million parameters.

  \begin{table}[htb]
    \begin{center}
    \caption{The embedding DNN architecture. x-vectors are extracted
    at layer $segment6$, before the nonlinearity. The $N$ in the softmax
    layer corresponds to the number of training languages.}
    \label{tab:xvect}
    \begin{tabular}{c|c|c|c}
    \hline
    Layer        &  Layer context         &   Total context & Input $\times$ output  \\
    \hline
    frame1       &  $[t - 2, t + 2]$      &  $5$   &  $200\times512$  \\
    frame2       &  $\{t - 2, t, t + 2\}$ &  $9$   &  $1536\times512$  \\
    frame3       &  $\{t - 3, t, t + 3\}$ &  $15$  &  $1536\times512$  \\
    frame4       &  $\{t\}$               &  $15$  &  $512\times512$  \\
    frame5       &  $\{t\}$               &  $15$  &  $512\times1500$  \\
    stats pooling&  $[0, T)$              &  $T$   &  $1500T\times3000$  \\
    segment6     &  $\{0\}$               &  $T$   &  $3000\times512$  \\
    segment7     &  $\{0\}$               &  $T$   &  $512\times512$  \\
    softmax      &  $\{0\}$               &  $T$   &  $512\times$$N$  \\
    \hline
    \end{tabular}
    \end{center}
    \end{table}

  We train the x-vector system with a combined dataset including
  AP16-OL7, AP17-OL3 and AP17-OLR-test, and the target number of the system
  refers to the number of all languages, i.e. $10$.
  This basic system is used for all three tasks with different back-end evaluation,
  either producing scores directly from the output of the original system for different target languages (task 1 and 2),
  or extracting x-vectors for each utterance for later identification (task 3).

  \subsection{Performance results}
  
  The primary evaluation metric in AP19-OLR is $C_{avg}$. Besides that, we also present the performance
  in terms of equal error rate (EER). These metrics evaluate
  system performance from different perspectives, offering a whole picture of the capability
  of the tested system.  The performance is evaluated on both the AP19-OLR-dev and AP19-OLR-test databases.
  Table~\ref{tab:results} shows the utterance-level C$_{avg}$ and EER results for the three tasks respectively.
  For task 1, we choose the short-utterance subset of AP18-OLR-test to be the development set.

  \begin{table}[htb]
    \begin{center}
    \caption{C$_{avg}$ and EER results of three tasks}
    \label{tab:results}
    \begin{tabular}{c|c|c|c|c}
    \hline
                & \multicolumn{2}{|c|}{Dev set} &  \multicolumn{2}{c}{Test set}\\
    \hline
    Task       &  $C_{avg}$  &   EER\% &  $C_{avg}$ &   EER\%  \\
    \hline
    short-utterance & 0.1271 &  12.37  &  0.1257   &  12.22      \\
    cross-channel   & 0.3868 &  43.13  &   0.3720  &  38.44     \\
    zero-resource   & 0.3393 & 34.47   &   0.2027  &  21.94     \\
    \hline
    \end{tabular}
    \end{center}
    \end{table}

  \subsubsection{Short-utterance LID}
  
  The first task identifies short-duration utterances.
  The test set is AP19-OLR-short which contains candidate speech segments with $1$ second duration.
  As the languages in AP19-OLR-short are the same as in the training set of our x-vector system,
  the output of the original system by propagating the test set can be seen as the 
  confidence that each speech segment is belonging to a specific language.
  From the `short-utterance' results in Table~\ref{tab:results}, we find that short-duration utterances
  are hard to recognize.

  \subsubsection{Cross-channel LID}
  The second task identifies six languages which are also included in the training set of above x-vector system,
  so the scores referring to those six languages for each utterance can also be produced as task 1 does.
  Cross-channel speech signals are much more difficult for the baseline system to recognize, 
  that can be seen from the `cross-channel' results in Table~\ref{tab:results}.



  \subsubsection{Zero-resource LID}
  The evaluation process for task 3 can be seen as identifying languages based on a reference dataset where 
  the languages may never be seen before.
  First we extract x-vectors for each segment in the reference set, 
  and then accumulate the utterance-level x-vectors for each language to produce
  language-level x-vectors. Each language-level x-vector can represent that specific language.
  We also extract x-vectors for each segment in the test set.
  Finally, we compare the utterance-level x-vectors from the test set to the language-level x-vectors from the 
  reference set respectively, then decide which language the test segments belong to.
  The metric used for the comparison in this paper is `cosine' distance.
  From the `zero-resource' results in Table~\ref{tab:results}, 
  it can be seen that resource-limited LID keeps challenging.
  The difference of target languages between development set and test set
  results in the gap of the performance.


  \section{Conclusions}
  
  We presented the data profile, task definitions and evaluation principles of the AP19-OLR challenge.
  To assist participants to construct a reasonable starting system, we published baseline
  system based on the x-vector model.
  We showed that the tasks defined by AP19-OLR are rather challenging and are worthy of careful study.
  All the data resources are free for the participants, and the recipes of the baseline systems can
  be freely downloaded.
  
  
  \section*{Acknowledgment}
  This work was supported in part by the National Natural Science
  Foundation of China under Projects 61633013.
  
  \bibliographystyle{IEEEtran}
  \bibliography{olr}

  \end{document}